\documentclass{svproc}
\usepackage{graphicx}%
\usepackage{multirow}%
\usepackage{amsmath,amssymb,amsfonts}%
\usepackage{mathrsfs}%
\usepackage[title]{appendix}%
\usepackage{xcolor}%
\usepackage{textcomp}%
\usepackage{manyfoot}%
\usepackage{booktabs}%
\usepackage{algorithm}%
\usepackage{algorithmicx}%
\usepackage{algpseudocode}%
\usepackage{listings}%
\usepackage{url}

\begin{document}
\mainmatter              
\title{Stochastic Degree Sequence Model with Edge Constraints (SDSM-EC) for Backbone Extraction}
\titlerunning{Stochastic Degree Sequence Model with Edge Constraints}  
%
\author{Zachary P. Neal\inst{1} \and Jennifer Watling Neal\inst{2}}
\authorrunning{Zachary P. Neal et al.} 
\institute{Michigan State University, East Lansing MI 48824, USA,\\
\email{zpneal@msu.edu.edu},\\ WWW home page:
\texttt{https://www.zacharyneal.com}
\and
Michigan State University, East Lansing MI 48824, USA}

\maketitle              

\begin{abstract}
It is common to use the projection of a bipartite network to measure a unipartite network of interest. For example, scientific collaboration networks are often measured using a co-authorship network, which is the projection of a bipartite author-paper network. Caution is required when interpreting the edge weights that appear in such projections. However, backbone models offer a solution by providing a formal statistical method for evaluating when an edge in a projection is statistically significantly strong. In this paper, we propose an extension to the existing Stochastic Degree Sequence Model (SDSM) that allows the null model to include edge constraints (EC) such as prohibited edges. We demonstrate the new SDSM-EC in toy data and empirical data on young children's' play interactions, illustrating how it correctly omits noisy edges from the backbone.

\\
\phantom{X}\\
\textbf{This is a post-print. For the version of record, please see:}\\
Neal, Z. P. and Neal, J. W. (2024). Stochastic Degree Sequence Model with Edge Constraints (SDSM-EC) for Backbone Extraction. Pp. 127–136 in \textit{Complex Networks \& Their Applications XII}, edited by Cherifi, H., Rocha, L. M., Cherifi, C., and Donduran, M. Springer. \url{https://doi.org/10.1007/978-3-031-53468-3_11}
\end{abstract}

\section{Introduction}  
It is common to use the projection of a bipartite network to measure a unipartite network of interest. For example, scientific collaboration networks are often measured using a co-authorship network, which is the projection of a bipartite author-paper network \cite{newman2001}. Similarly, corporate networks are often measured using a board co-membership or `interlocking directorate' network, which is the projection of a bipartite executive-board network \cite{burris2005interlocking}. The edges in a bipartite projection are weighted (e.g., number of co-authored papers, number of shared boards), but these weights do not provide an unbiased indicator the strength of the connection between vertices \cite{latapy2008,neal2014backbone}. To overcome this bias, backbone extraction identifies the edges that are stronger than expected under a relevant null model, retaining only these edges to yield a simpler unweighted network (i.e., the backbone) that is more suitable for visualization and analysis. 

Many null models exist for extracting the backbone of bipartite networks, with each model specifying different constraints on the random networks against which an observed network is compared. However, none of the existing models permit constraints on specific edges. In this paper, we extend the fastest and most robust existing backbone model -- the stochastic degree sequence model (SDSM) \cite{neal2021comparing} -- to accommodate one type of edge constraint: prohibited edges. Prohibited edges are edges that in principle cannot occur in the network, and can arise in many contexts. For example, in a bipartite author-paper network, an author cannot write a paper before their birth, and in a bipartite executive-board network, anti-trust laws prevent executives from serving on the boards of competitors. We illustrate the new stochastic degree sequence model with edge constraints (SDSM-EC) first in toy data, then in empirical data recording young childrens' membership in play groups.

\subsection{Preliminaries}
A bipartite network's vertices can be partitioned into two sets such that edges exist between, but not within, sets. In this work, we focus on a special case of a bipartite network -- a two-mode network -- where the two sets of vertices represent distinctly different entities that we call `agents' and `artifacts' (e.g. authors and papers, or executives and corporate boards). 

To facilitate notation, we represent networks as matrices. First, we represent a bipartite network containing $r$ `agents' and $c$ `artifacts' as an $r \times c$ binary incidence matrix $\mathbf{B}$, where $B_{ik} = 1$ if agent $i$ is connected to artifact $k$ (e.g., author $i$ wrote paper $k$), and otherwise is $0$. The row sums $R = r_1...r_c$ of $\mathbf{B}$ contain the degree sequence of the agents (e.g., the number of papers written by each author), while the column sums $C = c_1...c_r$ of $\mathbf{B}$ contain the degree sequence of the artifacts (e.g., the number of authors on each paper). A prohibited edge in a bipartite network is represented by constraining a cell to equal zero, and therefore is sometimes called a `structural zero' \cite{rsiena}. Second, we represent the projection of a bipartite network as an $r \times r$ weighted adjacency matrix $\mathbf{P} = \mathbf{BB}^T$, where $\mathbf{B}^T$ represents the transpose of $\mathbf{B}$. In $\mathbf{P}$, $P_{ij}$ equals the number of artifacts $k$ that are adjacent to both agent $i$ and agent $j$ (e.g., the number of papers co-authored by authors $i$ and $j$). Finally, we represent the backbone of a projection $\mathbf{P}^\prime$ as an $r \times r$ binary adjacency matrix, where $P^\prime_{ij} = 1$ if agent $i$ is connected to agent $j$ in the backbone, and otherwise is 0.

Let $\mathcal{B}$ be an ensemble of $r \times c$ binary incidence matrices, which can be constrained to have certain features present in $\mathbf{B}$. Let $P^*_{ij}$ be a random variable equal to $(\mathbf{B^*}\mathbf{B^*}^T)_{ij}$ for $\mathbf{B^*}~\in~\mathcal{B}$. Decisions about which edges appear in a backbone extracted at the statistical significance level $\alpha$ are made by comparing $P_{ij}$ to $P^*_{ij}$:
\[ P_{ij}'=
\begin{cases}
1 & \text{ if } \Pr(P^*_{ij} \geq P_{ij}) < \frac{\alpha}{2},\\
0 & \text{otherwise.}
\end{cases}\]
This test includes edge $P'_{ij}$ in the backbone if its weight in the observed projection $P_{ij}$ is uncommonly large compared to its weight in projections of members of the ensemble $P^*_{ij}$.

\section{Backbone models}
\subsection{The stochastic degree sequence model (SDSM)}
Models for extracting the backbone of bipartite projections differ in the constraints they impose on $\mathcal{B}$. The most stringent model -- the Fixed Degree Sequence Model (FDSM) \cite{zweig2011systematic} -- relies on a microcanonical ensemble that constrains each member of $\mathcal{B}$ to have exactly the same row and column sums as $\mathcal{B}$. Computing $P^*_{ij}$ under the FDSM is slow because it requires approximation via computationally intensive Monte Carlo simulation. Despite recent advances in the efficiency of these simulations \cite{godard2022fastball}, it is often more practical to use the less stringent Stochastic Degree Sequence Model (SDSM) \cite{neal2014backbone}. The SDSM relies on a canonical ensemble that constrains each member of $\mathcal{B}$ to have the same row and column sums as $\mathbf{B}$ \textit{on average}. SDSM is fast and exact, and comparisons with FDSM reveal that it yields similar backbones \cite{neal2021comparing}.

Under the SDSM, $P^*_{ij}$ follows a Poisson-binomial distribution whose parameters can be computed from the entries of probability matrix $\mathbf{Q}$, where $Q_{ik} = \Pr(B^*_{ik} = 1) \text{ for } \mathbf{B^*}~\in$ a microcanonical $\mathcal{B}$. That is, $Q_{ik}$ is the probability that $B^*_{ik}$ contains a $1$ in the space of all matrices with given row and column sums. Most implementations of SDSM approximate $\mathbf{Q}$ using the fast and precise Bipartite Configuration Model (BiCM) \cite{saracco2015randomizing,saracco2017inferring}. However, it can also be computed with minimal loss of speed and precision \cite{neal2021comparing} using a logistic regression \cite{neal2014backbone}, which offers more flexibility. This method estimates the $\beta$ coefficients in $$B_{ik} = \beta_0 + \beta_1r_i + \beta_2c_k + \epsilon$$ using maximum likelihood, then defines $Q_{ik}$ as the predicted probability that $B_{ik} = 1$.

\subsection{The stochastic degree sequence model with edge constraints (SDSM-EC)}
The constraints that SDSM imposes on $\mathcal{B}$ are determined by the way that $\mathbf{Q}$ is defined. In the conventional SDSM, $\mathbf{Q}$ is defined such that $Q_{ik}$ is the probability that $B^*_{ik}$ contains a $1$ in the space of all matrices with given row and column sums, which only imposes constraints on the row and column sums of members of $\mathcal{B}$. To accommodate edge constraints, we define $\mathbf{Q}^\prime$ such that $Q^\prime_{ik}$ is the probability that $B^*_{ik}$ contains a $1$ in the space of all matrices with given row and column sums \textit{and no 1s in prohibited cells}.

The BiCM method cannot be used to approximate $\mathbf{Q}^\prime$, however the logistic regression method can be adapted to approximate it. If $B_{ik}$ is a prohibited edge, then $Q_{ik} = 0$ by definition. If $B_{ik}$ is not a prohibited edge, then $Q_{ik}$ is the predicted probability that $B_{ik} = 1$ based on a fitted logistic regression. Importantly, however, whereas the logistic regression used to estimate $\mathbf{Q}$ is fitted over all $B_{ik}$, the logistic regression used to estimate $\mathbf{Q}^\prime$ is fitted only over $B_{ik}$ that are not prohibited edges.

\section{Results}
\subsection{Estimating $\mathbf{Q}^\prime$}
In general the true values of $Q_{ik}$ are unknown. However, for small matrices they can be computed from a complete enumeration of the space. To evaluate the precision of $Q_{ik}$ estimated using the SDSM-EC method described above, we first enumerated all $4 \times 4$ incidence matrices with row sums \{1,1,2,2\} and column sums \{1,1,2,2\}; there are 211. Next, we constrained this space to matrices in which a randomly selected one or two cells always contain a zero (i.e. bipartite networks with one or two prohibited edges). Finally, we computed the true value of each $Q_{ik}$ for all cells and all spaces, estimated each $Q_{ik}$ using the logistic regression method, and computed the absolute deviation between the two.

\begin{figure}
\centering
\includegraphics[width=\textwidth]{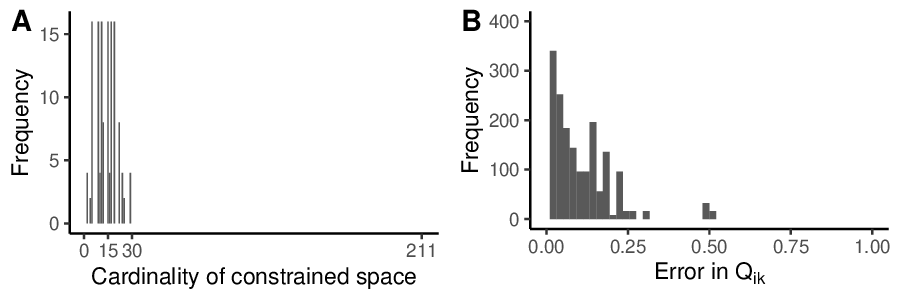}
\caption{(A) The cardinality of the space of matrices with row sums \{1,1,2,2\} and column sums \{1,1,2,2\} and one or two cells constrained to zero is small compared to the cardinality of the space without constrained cells. (B) The deviation between the true and estimated $Q_{ik}$ for all such constrained spaces tends to be small.}
\label{fig:precision}
\end{figure}

Figure \ref{fig:precision}A illustrates that, compared to the cardinality of the unconstrained space ($|\mathcal{B}| = 211$), the cardinalities of the spaces constrained by one or two prohibited edges are much lower ($|\mathcal{B}| = 2 - 29$, gray bars). That is, while the SDSM evaluates whether a given edge's weight is significant by comparing its value to a large number of possible worlds, the SDSM-EC compares its value to a much smaller number of possible worlds. Figure \ref{fig:precision}B illustrates the deviations between the true value of $Q_{ik}$ and the value estimated using the logistic regression method. It demonstrates that although SDSM-EC requires approximating $Q_{ik}$, these approximations tend to be close to the true values.

\subsection{Toy illustration}
\begin{figure}
\centering
\includegraphics[width=\textwidth]{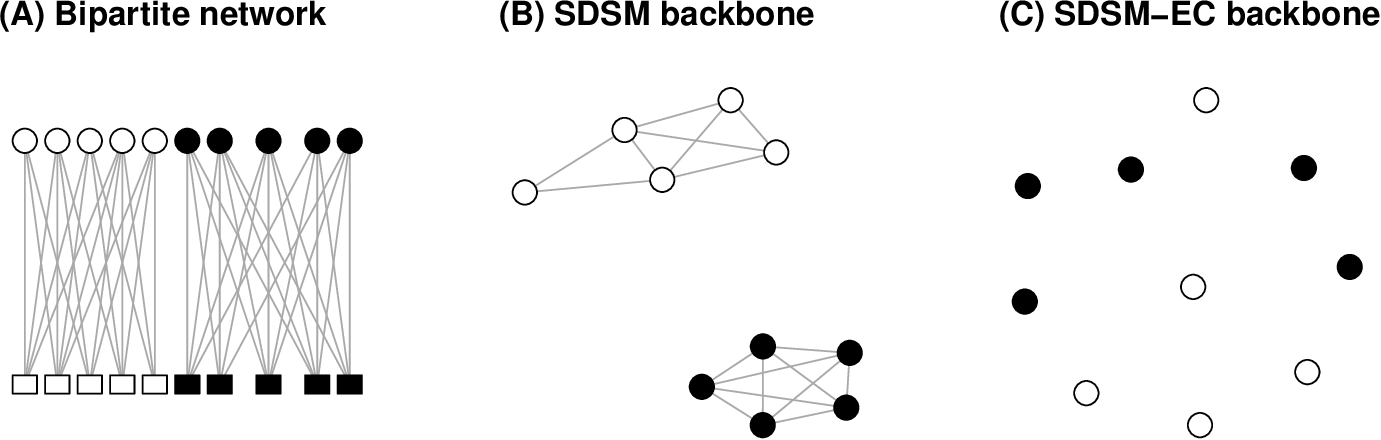}
\caption{(A) A bipartite network containing two groups of agents and two groups of artifacts, such that agents are connected only to their own group's artifacts. (B) The SDSM backbone of a projection of this bipartite graph, which assumes that an agent \textit{could} be connected to another group's artifact, suggests within-group cohesion among agents. (C) The SDSM-EC projection, which assumes that an agent \textit{could not} be connected to another group's artifact, suggests none of the edges in the projection are significant.}
\label{fig:toy}
\end{figure}

We offer a toy example to illustrate the impact of imposing edge constraints in backbone extraction. Figure \ref{fig:toy}A illustrates a bipartite network that contains two types of agents (open and filled circles) and two types of artifacts (open and filled squares), such that agents are only connected to artifacts of the same type. Such a network might arise in the context of university students joining clubs. For example, suppose Harvard students (open circles) only join Harvard clubs (open squares), while Yale students (filled circles) only join Yale clubs (filled squares).

Figure \ref{fig:toy}B illustrates the backbone extracted from a projection of this bipartite network using the SDSM. Using the SDSM implies that there are no constraints on edges in the null model. In the context of student clubs, this means that in the null model it is possible for a Harvard student to join a Yale club, and vice versa, and that the pattern of segregation that appears in the bipartite network is chosen (i.e. homophily). The SDSM backbone displays a high level of within-group cohesion (i.e. homophily). This occurs for two reasons. First, agents from the same group share many artifacts (e.g., two Harvard students belong to many of the same clubs). Second, if agents were connected to artifacts randomly (e.g., Harvard students joined both Harvard and Yale clubs), as the SDSM null model assumes, then agents from the same group would have shared fewer artifacts. The presence of within-group connections in the SDSM backbone reflects the fact that it is noteworthy that pairs of Harvard students, or pairs of Yale students, are members of many of the same clubs because they could have chosen otherwise.

Figure \ref{fig:toy}C illustrates the backbone extracted using the SDSM-EC, where we specify that edges are prohibited between an agent and artifact of a different type. In the context of student clubs, this means that in the null model it is \textit{not} possible for a Harvard student to join a Yale club, and vice versa, and that the pattern of segregation is enforced by university regulations. The SDSM-EC backbone is empty. This occurs because although agents from the same group share many artifacts, they also share many artifacts under the null model. The absence of connections in the SDSM-EC backbone reflects the fact that it is uninteresting that pairs of Harvard students, or pairs of Yale students, are members of many of the same clubs because they could not have chosen otherwise.

\subsection{Empirical illustration}
We offer an empirical example of the application of SDSM-EC to illustrate its practicality and impact. It can be difficult to directly measure social networks among very young children. One alternative is to infer these networks from observations of their play groups using bipartite backbones \cite{neal2022inferring}. However, considering edge constraints can be important because the organization of the school can mean that it may be impossible to observe certain children playing together.

These data were collected in Spring 2013 by observing the behaviors of 53 children in a preschool in the Midwestern United States \cite{gornik2018connections,neal2022role,neal2017codevelopment,neal2022inferring}. A scan observation method was employed whereby a randomly selected child was observed for a period of 10 seconds. After the 10 second period had elapsed, the trained observer coded the child's predominant behavior and, if applicable, the peers with whom they were interacting \cite{hanish2005exposure}. Here, we focus only on social play behaviors because they were the most common form of social behavior, and the most likely to involve direct interaction with peers. A total of 1829 social play events were observed during data collection. These data are organized as a bipartite network $\mathbf{B}$ where $B_{ik} = 1$ if child $i$ was observed participating in a play group during observation $k$. A projection of $\mathbf{P} = \mathbf{BB}^T$, where $P_{ij}$ indicates the number of times children $i$ and $j$ were observed playing together provides an indirect indicator of the children's' social network, particularly when refined using backbone extraction \cite{neal2022inferring}.

In this context, two types of prohibited edges exist in the bipartite network. First, the school was organized into two age-based classrooms, a classroom of 3-year-olds and a classroom of 4-year-olds. Because these classrooms used different spaces, it was not possible to observe a 3-year old and a 4-year-old together. Therefore, edges from 3-year-olds to observations of 4-year-olds are prohibited, and likewise edges from 4-year-olds to observations of 3-year-olds are prohibited. Second, the children varied in their attendance status: some attended for the full day, some attended only in the morning, and some attended only in the afternoon. Because attendance status determines which children were present and able to play together, it was not possible to observe an AM child and a PM child together. Therefore, edges from AM children to observations conducted in the afternoon are prohibited, and likewise edges from PM children to observations conducted in the morning are prohibited.

\begin{figure}
\centering
\includegraphics[width=\textwidth]{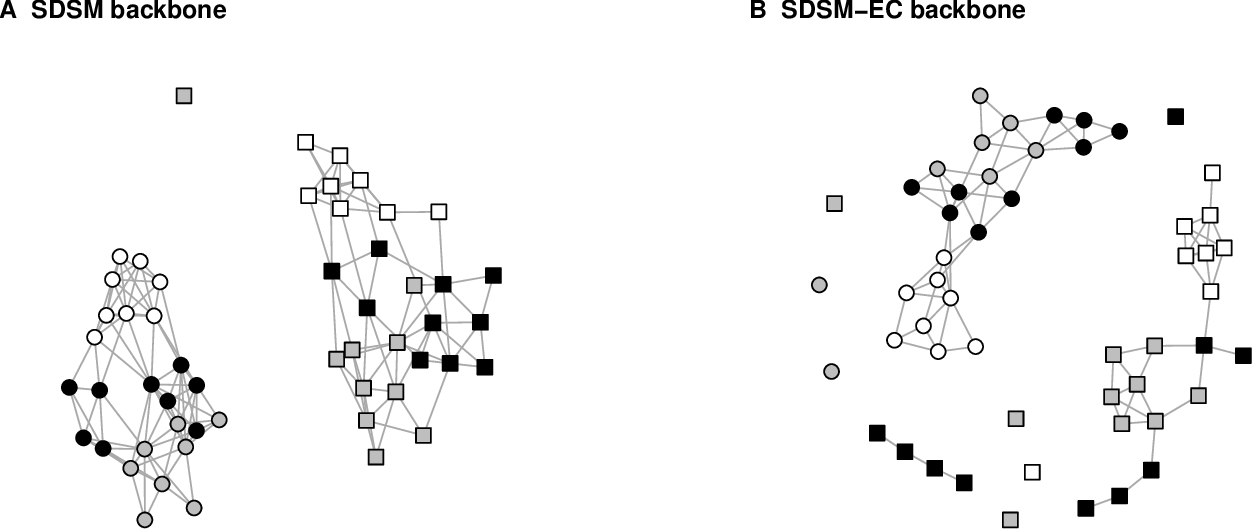}
\caption{(A) Backbone extracted using SDSM and (B) SDSM-EC from 1829 observations of 53 preschool childrens' play groups. Vertex shape represents age-based classrooms: circles = 3 year old classroom, squares = 4 year old classroom. Vertex color represents attendance status: black = full day, gray = AM only, white = PM only.}
\label{fig:empirical}
\end{figure}

Figure \ref{fig:empirical} illustrates two backbones extracted from these data, using shape to represent classroom (circles = 3-year-olds, squares = 4-year-olds) and color to represent attendance status (black = full day, gray = AM only, white = PM only). Figure \ref{fig:empirical}A was extracted using the SDSM and therefore does not consider these edge constraints, while Figure \ref{fig:empirical}B was extracted using the SDSM-EC and does consider these edge constraints. There are some similarities between the SDSM and SDSM-EC backbones that reflect characteristics of the setting: 3-year-olds (circles) are never connected to 4-year-olds (squares), and AM children (gray) are never connected to PM children (white), because it was not possible to observe such children together. However, there are also differences that highlight the impact of incorporating edge constraints using SDSM-EC. The SDSM-EC backbone contains many fewer edges ($E = 85$) than the SDSM backbone ($E = 153$). This occurs for similar reasons to the loss of edges in the toy example above, although is less extreme. 

A hypothetical example serves to illustrate why the SDSM-EC backbone contains fewer edges in this context. Consider the case of an AM child and a Full Day child in the 3-year-old classroom who were observed to play together a few times. The SDSM compares this observed co-occurrence to the expected number of co-occurrences if these two children had played with other AM or Full Day children and with others in the 3-year-old classroom (which is possible), but also if they had played with PM children and children in the 4-year-old classroom (which is not possible). Under such a broad null model that includes some impossible play configurations, observing these two children playing together even just a few times seems noteworthy, and therefore an edge between them is included in the backbone. In contrast, the SDSM-EC compares this observed co-occurrence to the expected number of co-occurrences if these two children had played with other AM or Full Day children and with others in the 3-year-old classroom only, recognizing that it was not possible for the AM child to play with PM children or for either to play with children in the 4-year-old classroom. Under this more constrained null model that excludes impossible play configurations, observing these two children playing together just a few times is not particularly noteworthy, and therefore an edge between them is omitted from the backbone. As this example illustrates, the SDSM-EC contains fewer edges because it correctly omits edges that might seem significantly strong when evaluated against a null model that includes impossible configuration, but that are not significantly strong when evaluated against a properly constrained null model that excludes impossible configurations.

\section{Conclusion}
Although bipartite projections offer a promising way to indirectly measure unipartite networks of interest, caution is required when interpreting the edge weights that appear in such projections. Backbone models offer a solution by providing a formal statistical method for evaluating when an edge in a projection is statistically significantly strong by comparison to a bipartite null model. However, extracting an accurate backbone using these methods requires that the null model is properly constrained. In many cases the FDSM (slower) and SDSM (faster) are appropriate and yield similar results \cite{neal2021comparing}, however these null models only constrain the degree sequences, but cannot impose edge constraints such as prohibited edges.

In this work, we have introduced the SDSM-EC, an extension of SDSM that allows the user to specify edge constraints in the form of prohibited edges. Prohibited edges arise in bipartite networks when a given agent cannot be connected to a given artifact, for example, because the agent is not present or because such a connection is legally prohibited. We have demonstrated in both a toy example and an empirical example that the SDSM-EC performs as expected, correctly omitting weaker edges in the backbone that are not significant when these constraints are considered, but that might have erroneously appeared significant under the SDSM. Therefore, we recommend that SDSM-EC be used to extract the backbones of bipartite projections when the bipartite network contains prohibited edges. The SDSM-EC is implemented in the \texttt{sdsm()} function of the \texttt{backbone} package for R \cite{neal2022backbone}.

We have focused on one common type of edge constraint: prohibited edges. However, a second type of edge constraint is also possible: required edges. Required edges arise in bipartite networks when a given agent must always be connected to a given artifact, for example, because the agent is the initiator of the artifact (e.g. a paper's lead author, a club's founder). It is trivial to extend the SDSM-EC to also accommodate such constraints. When $\mathbf{Q}$ is estimated, $Q_{ik} = 0$ for prohibited edges and $Q_{ik} = 1$ for required edges, then the remaining $Q_{ik}$ values are computed using the same logistic regression method described above.

This work highlights the importance of using a properly constrained null model when extracting the backbone of bipartite projections, and identifies several avenues for future research. First, while $\mathbf{Q}$ under the SDSM can be estimated quickly and precisely using the BiCM \cite{saracco2015randomizing,saracco2017inferring}, $\mathbf{Q}$ under the SDSM-EC must be estimated using logistic regression, which is slower and less precise \cite{neal2021comparing}. Future work should investigate improved methods for estimating $\mathbf{Q}$, which has the potential to benefit not only the SDSM-EC, but all variants of the SDSM. Second, while a broad class of bipartite null models exist \cite{strona2018bi} and now include edge constraints, future work should investigate the importance and feasibility of incorporating other types of constraints. \\

\noindent\textbf{Acknowledgements.} We thank Emily Durbin for her assistance collecting the empirical data.\\

\noindent\textbf{Data availability statement.} The data and code necessary to reproduce the results reported above are available at \url{https://osf.io/7z4gu}.

\end{document}